\documentclass[showpacs,prb,twocolumn,groupedaddress,floatfix]{revtex4-1}
\usepackage{bm}
\usepackage{pstricks}
\usepackage{amsmath,amssymb,amsbsy}
\usepackage{graphicx,subfigure}
\usepackage{verbatim}
\usepackage[utf8]{inputenc}
\usepackage[USenglish]{babel}
\usepackage[T1]{fontenc}
\usepackage{graphicx}
\usepackage{dcolumn}
\usepackage{multirow}
\usepackage{latexsym}
\usepackage{tikz}
\usepackage{pgf}
\usepackage{pbox}
\usepackage{float}
\usepackage{multirow}
\usepackage{braket}
\usepackage{booktabs}
\usepackage{hyperref}

\usepackage{xcolor}
\hypersetup{
    colorlinks,
    linkcolor={blue!50!black},
    citecolor={blue!50!black},
    urlcolor={blue!80!black}
}

\begin{document}

\title{Improvements in the $GW$ and BSE calculations on phosphorene}

\author{F. Ferreira}

\affiliation{Centro de F\'{i}sica and Departamento de F\'{i}sica and QuantaLab, Universidade
do Minho, Campus de Gualtar, Braga 4710-057, Portugal}

\author{R.~M. Ribeiro}

\affiliation{Centro de F\'{i}sica and Departamento de F\'{i}sica and QuantaLab, Universidade
do Minho, Campus de Gualtar, Braga 4710-057, Portugal}

\date{\today}

\begin{abstract}
Phosphorene is a bidimensional material that has properties useful for semiconductor devices.
In this work we studied the electronic and optical properties of this material using the $GW$ approximation and the Bethe-Salpeter equation (BSE) methods.
We stress the importance of a careful convergence study of the most relevant parameters,
and we show how they affect the result of the calculations.
A comparison with previous results is given.
The QP band gap obtained was 2.06 eV  and it is in good agreement with experimental results.
BSE calculations were performed on top of $G_0W_0$ in order to include  excitonic effects.
The absorption spectrum was analyzed and an optical gap of 1.22 eV was obtained.
The calculated excitonic binding energy is 0.84 eV, also in good agreement with experimental results.

\end{abstract}

\maketitle
\section{Introduction}

\begin{figure}[]
\centering
\includegraphics[scale=0.1]{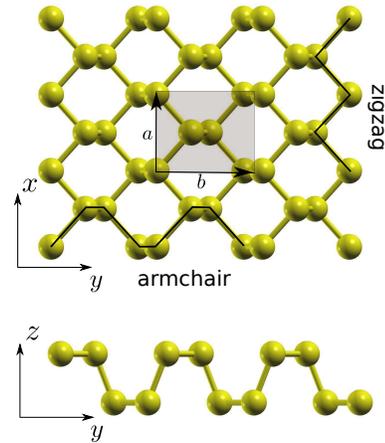}
\caption[Structure of phosphorene.]
{(Color online) Structure of phosphorene. Phosphorus atoms are in yellow.
Top: top view in $xy$ plane. Bottom: side view in $zy$ plane.
Phosphorene has a zigzag structure along the $x$ axis and an armchair structure along the $y$ axis. }
\label{P_structure}
\end{figure}

Since its recent synthesis\cite{doi:10.1021/nl502892t}, phosphorene has been attracting interest due to its peculiar and distinct properties.
Unlike graphene, this material is a semiconductor with a band gap of approximately 2.0 eV,\cite{doi:10.1021/nl502892t} which is much higher than
the bulk band gap of 0.3 eV.\cite{PhysRev.92.580}
A wide band gap makes it very useful for semiconductor applications.
Studies have reported high mobility\cite{highmobph} and high current on-off ratio for field effect transistors.\cite{doi:10.1021/nn501226z}
Phosphorene can be stacked in two or more layers due to the weak Van der Walls interlayer interactions.\cite{Li2014}
It was shown that the band gap value can be tuned by the number of stacked layers.\cite{PhysRevB.89.235319}
This stacking can be used to control the electronic and optical properties by changing the number of the stacked layers.
The infrared properties of Phosphorene have been recently measured\cite{ncomms14071} and found to depend on uniaxial strain and useful for infrared photonics.
Band alignment on heterostructures involving phosphorene has been theoretically studied\cite{PhysRevB.94.035125,APS2016} and
a high power conversion efficiency for solar cells was found.\cite{APS2016}
Another interesting property is the strong in-plane anisotropy of transport and optical properties.
This is due to its anisotropic crystal structure, shown in Figure \ref{P_structure}.
Phosphorene has a puckered lattice due to sp$^3$ hybridization,
which also allows band gap engineering by applying strain in-plane or out-plane.\cite{PhysRevB.90.205421}
It was also shown that the deformation of phosphorene in a specific direction can be used to induce a semiconductor-metal
transition.\cite{doi:10.1021/nl500935z}
Phosphorene was shown to have negative Poisson's ratio\cite{Jiang2014} and mechanical flexibility\cite{doi:10.1063/1.4885215} that allows
its use under extreme mechanical conditions.
Also it can be used for gas sensors\cite{doi:10.1021/jz501188k} and solar cell applications.\cite{doi:10.1021/nn5027388}

Density Functional Theory (DFT) has been extensively used to calculate electronic and optical properties.
Yet, it is well known that DFT, in the most common exchange-correlation approximations used,
does not provide a good description of the electronic band structures, namely the band gap is underestimated.
Also, because phosphorene has low dimensionality, the excitonic effects can not be neglected and must be taken into account
if we want to study its optical properties.
DFT does not include excitonic effects.

In this paper we study the electronic properties of phosphorene using the $GW$ quasiparticle (QP) approximation.\cite{PhysRev.139.A796,HedinS}
To calculate the $GW$ spectrum we have to determine the electron self-energy operator $\Sigma$, which is equivalent to solve a many-body problem.
The self-energy operator is approximated and expressed
in terms of the screened Coulomb potential $W$ as $\Sigma \approx i GW$, with $G$ being the single-particle Green's function.
To include the excitonic effects and obtain the optical properties we use a two-particle Green's function formalism,
through the Bethe-Salpeter equation (BSE)\cite{PhysRev.84.1232, PhysRevLett.80.4510} on top of the $GW$ approximation.

Phosphorene has been studied before using the $GW$ approximation\cite{doi:10.1021/nl502892t, PhysRevB.92.085419, PhysRevB.89.201408}
and also the $GW+$BSE method,\cite{PhysRevB.89.235319, PhysRevB.90.205421,2053-1583-2-4-044014}
but the results of these works do not agree (see tables \ref{table:GW_summary_P} and \ref{table:BSE_P_summary}).
One possible reason may be different criteria for convergence.
Ref. \onlinecite{PhysRevB.90.205421} calculates $G_0W_0$ with 112 and 326 empty bands and a vacuum size of 28.34 bohr, while
Ref. \onlinecite{PhysRevB.92.085419} and \onlinecite{PhysRevB.89.201408} used 360 bands (90 bands per atom).
Ref. \onlinecite{PhysRevB.89.235319} used 10 times the valence bands to converge the dielectric matrix.
These numbers of bands are small when compared to the ones that we found good for convergence in our calculations.
Ref. \onlinecite{2053-1583-2-4-044014} does not mention the number of bands used for the calculation.
The truncation technique, which is needed to avoid interactions with periodic images, is not used in some of these works.\cite{doi:10.1021/nl502892t,
PhysRevB.90.205421, PhysRevB.92.085419, PhysRevB.89.201408}


\begin{table}[t]
\centering
\caption
{Summarized results for different works that used $GW$ approximation, including this work with a vacuum distance of 50 bohr.
$L$ stands for the vacuum size between the layers (see main text).
$GW_0$  is an update of the  $G_0W_0$ calculation in which only $G$ is updated by converging the QP energies.}
\begin{ruledtabular}
\begin{tabular*}{\columnwidth}{@{\extracolsep{\fill} }lccr}
Reference                            & Type of calculation & $L$ [bohr] & Band gap [eV]  \\ \midrule
   This Work                               &      $G_0W_0$ &  50 &  2.06 \\
   Ref. \onlinecite{doi:10.1021/nl502892t} &      $GW_0$   &  34 &  1.94 \\
   Ref.  \onlinecite{PhysRevB.89.235319}   &      $G_0W_0$ &   - &  2.00 \\
   Ref. \onlinecite{PhysRevB.90.205421}    &      $G_0W_0$ &  28 &  2.31 \\
   Ref.  \onlinecite{PhysRevB.92.085419}   &      $GW_0$   &  28 &  1.85 \\
   Ref.  \onlinecite{PhysRevB.89.201408}   &      $G_0W_0$ &  38 &  1.60 \\
   Ref. \onlinecite{2053-1583-2-4-044014}  &      $G_0W_0$ &   - &  2.00 \\
   Ref. \onlinecite{doi:10.1021/nl502892t} &  Experimental &   - &  2.05 \\
   Ref. \onlinecite{natureHighly}          &  Experimental &   - &  $2.2 \pm 0.1$   \\
 \end{tabular*}
\end{ruledtabular}
\label{table:GW_summary_P}
\end{table}


Concerning the BSE results, we found that we have to use a grid of $90 \times 90 \times 1$ $\mathbf{k}$-points in order
to achieve convergence (see subsection \ref{ssec:BSE} below), a value larger than some of the works published.\cite{PhysRevB.90.205421,PhysRevB.89.235319,2053-1583-2-4-044014}


\begin{table}[b]
\centering
\caption[Summarized results for different works that use the BSE on phosphorene]
{Summarized results for different works that use the BSE on phosphorene.
EBE stands for excitonic binding energy.}
\begin{ruledtabular}
\begin{tabular*}{\columnwidth}{@{\extracolsep{\fill} }lccr}
 Reference      & Type of calculation & Optical gap [eV]  & EBE [eV]   \\ \midrule
 This Work      &    $G_0W_0+$BSE          &  1.22      &  0.84 \\
 Ref. \onlinecite{PhysRevB.89.235319} &     $G_0W_0+$BSE         &   1.20          & 0.80     \\
 Ref. \onlinecite{PhysRevB.90.205421}  &     $G_0W_0+$BSE        &  1.61          & 0.70    \\
 Ref. \onlinecite{2053-1583-2-4-044014} &  $G_0W_0+$BSE & 1.24 & 0.78 \\
 Ref. \onlinecite{natureHighly} & Experimental & $1.3 \pm 0.02$   &  $0.9 \pm 0.12$ \\
\end{tabular*}
\end{ruledtabular}
\label{table:BSE_P_summary}
\end{table}

In this work we study phosphorene properties with $GW+$BSE, since to our knowledge very few works have been done.
We also expect to contribute to clarify the convergence studies.
The $GW+$BSE approach is much more computationally expensive than DFT,
because of the interdependence of the convergence parameters and higher sensitivity to
the number of $\mathbf{k}$-points and number of bands, as will be explained in Sec. \ref{sec:method}.
A common error may be to not consider that convergence parameters are interdependent
and so can result in a false convergence behavior.
For instance, Ref. \onlinecite{PhysRevLett.105.146401}, shows that one of the causes of underestimation
of the QP band gap of ZnO was the result of an inadequate treatment of the convergence parameters.
The demanding resources that are needed to obtain rigorous results in $GW+$BSE calculations may enhance this problem.

This paper is organized as follows.
In Sec. \ref{sec:method}, the computational details are described, with an emphasis on the parameters that may affect convergence.
Section \ref{sec:results} includes the results of the calculations done with the $GW$ approximation and the BSE and the discussion.
The conclusions are given in Sec. \ref{sec:conclusion}.

\section{Method}
\label{sec:method}

This work was done in three steps. In the first step, DFT calculations were performed
using the open source software package {\sc Quantum ESPRESSO}.\cite{QE-2009}
In the second step, single-shot $GW$ ($G_0W_0$) calculations were done using the DFT results.
Lastly, BSE was applied on $G_0W_0$ calculations.
Both the $G_0W_0$ and BSE calculations were performed with the software package {\sc BerkeleyGW}.\cite{Deslippe20121269}

\subsection{DFT calculations}

DFT calculations were done using a scalar-relativistic norm-conserving pseudopotential.
The exchange-correlation functional used was the generalized gradient approximation of Perdew-Burke-Ernzerhof~(GGA-PBE).\cite{PhysRevLett.77.3865}
The plane-wave cut-off used was 70 Ry and the integration over the Brillouin-zone
was performed using the scheme proposed by Monkhorst-Pack~\cite{PhysRevB.13.5188} with a grid of
$9\times 9\times 1$ $\textbf{k}$-points.
Both the plane-wave cut-off and the $\textbf{k}$-points grid were chosen after a convergence analysis.
A vacuum size ($L$) between the layer images equal or greater than 20~bohr was enough to avoid interactions between the periodic images.

\subsection{$G_0W_0$ calculations}

$G_0W_0$ calculations were performed on top of DFT.
We used the Generalized-Plasmon-Pole (GPP) model  proposed by  Hybertson and Louie,\cite{PhysRevB.34.5390}
where we compute the static dielectric matrix  and then  extend it to finite frequencies.
This is done by considering the static polarizability matrix
\begin{multline}
\chi_{\mathbf{GG'}}(\mathbf{q},0)  = \sum_{\mathbf{k}}\sum_n^{occ} \sum_{n'}^{emp} M_{nn'}^*(\mathbf{k},\mathbf{q},\mathbf{G})
M_{nn'}(\mathbf{k},\mathbf{q},\mathbf{G}') \label{APBGW_static_POL} \\
 \times \frac{1}{E_{n \mathbf{k}+\mathbf{q}} - E_{n' \mathbf{k}}} ,
\end{multline}
where
\begin{equation}
M_{nn'}(\mathbf{k},\mathbf{q},\mathbf{G}) = \bra{n \mathbf{k}+\mathbf{q}}
e^{i(\mathbf{q}+\mathbf{G}).\mathbf{r}} \ket{n' \mathbf{k}}
\label{APBGW_matrix_elements}
\end{equation}
are the plane-wave matrix elements,
$\mathbf{q} =\mathbf{k} - \mathbf{k}'$ is the momentum transfer, $\mathbf{G}$ is the reciprocal lattice vector,
$n$ and $n'$ are the band numbers of occupied ($occ$) and empty ($emp$) bands respectively,
$\ket{n, \mathbf{k}}$ and $E_{n, \mathbf{k}}$ are the DFT eigenvectors and eigenvalues respectively.
The plane-wave matrix elements are evaluated up to $\left| \mathbf{q}+\mathbf{G}  \right|^2$,
where $\left| \mathbf{q}+\mathbf{G}'  \right|^2 < E_{cut}$ defines the dielectric matrix cut-off $E_{cut}$.
The dielectric matrix is given by
\begin{equation}
\epsilon_{\mathbf{GG'}}(\mathbf{q},0) = \delta_{\mathbf{GG'}} - v(\mathbf{q}+\mathbf{G})
\chi_{\mathbf{GG'}}(\mathbf{q},0),
\label{APBGW_epsilon_matrix_satic}
\end{equation}
where $v(\mathbf{q}+\mathbf{G}) = \frac{4 \pi}{ \left| \mathbf{q}+\mathbf{G} \right|^2}$ is the bare Coulomb potential.
To construct $\epsilon_{\mathbf{GG'}}$ we have to do a summation on the empty bands and on the $\mathbf{G}$ vectors
which are given by the  $E_{cut}$.
These parameters have to be consistent with each other:
if we choose for instance 100 empty bands for the summation, then the energy of the 100th empty band is the $E_{cut}$ energy.
So we have just one convergence parameter for the construction of the $\epsilon_{\mathbf{GG'}}$: either $E_{cut}$ or the number of empty bands used.
Here we will usually use the number of empty bands for this convergence criterion.
The GPP model allows us to extend the static dielectric matrix to a finite frequency dependent matrix and then calculate $\Sigma$.
It is divided in two terms $\Sigma = \Sigma_{SX} + \Sigma_{CH}$, where $\Sigma_{SX}$ is the screened-exchange term and
the $\Sigma_{CH}$ is the Coulomb-hole term and they have the following expression
\begin{multline}
\bra{n\mathbf{k}} \Sigma_{SX} (\omega)  \ket{n'\mathbf{k}} \\ = - \sum_{n''}^{occ} \sum_{\mathbf{q}
\mathbf{GG}'}
M_{n''n}^*(\mathbf{k},\mathbf{-q},\mathbf{-G}) M_{n'' n'} (\mathbf{k},\mathbf{-q},\mathbf{-G}') \\
\times \left [  \delta_{\mathbf{GG'}} + \frac{\Omega_{\mathbf{GG}'}^{2}(\mathbf{q}) (1-i \
\tan \phi_{\mathbf{GG'}} (\mathbf{q})) }
{(\omega-E_{n'' \mathbf{k}-\mathbf{q}})^2 - \tilde{\omega}_{\mathbf{GG'}}^2(\mathbf{q})}    \right]
v (\mathbf{q}+\mathbf{G}'),
\label{BGW_SX_PPA}
\end{multline}

\begin{multline}
\bra{n\mathbf{k}} \Sigma_{CH} (\omega)  \ket{n'\mathbf{k}} \\ =  \frac{1}{2  }\sum_{n''} \sum_{\mathbf{q}\mathbf{G}\mathbf{G}'}
M_{n''n}^*(\mathbf{k},\mathbf{-q},\mathbf{-G}) M_{n'' n'} (\mathbf{k},\mathbf{-q},\mathbf{-G}') \\
\times
\frac{\Omega_{\mathbf{G}\mathbf{G}'}^{2}(\mathbf{q}) (1-i \   \tan \phi_{\mathbf{G}\mathbf{G}'} (\mathbf{q})) }
{\tilde{\omega}_{\mathbf{GG'}}(\mathbf{q})(\omega-E_{n'' \mathbf{k-q}} -  \tilde{\omega}_{\mathbf{GG'}}(\mathbf{q}))}
v (\mathbf{q}+\mathbf{G}').
\label{BGW_CH_PPA}
\end{multline}
The $\Omega_{\textbf{G}\textbf{G}'}$  is the effective bare plasma frequency and it is given by
\begin{equation}
\Omega_{\textbf{G}\textbf{G}'}^2(\mathbf{q}) = \omega_{p}^2 \frac{(\textbf{q} +
\textbf{G})\cdot (\textbf{q}+\textbf{G}')}
{\left| \textbf{q}+\textbf{G}\right|^2}\frac{\rho(\textbf{G}-\textbf{G}')}{\rho(0)} ,
\label{PPA_fsum}
\end{equation}
where $\rho$ is the electron charge density in reciprocal space and $\omega_p^2$ is the classical plasma frequency
defined by $\omega_p^2 = 4 \pi \rho (0) e^2/m$.
The $\tilde{\omega}_{\textbf{G}\textbf{G}'}$ is the GPP mode frequency and is related to $\Omega_{\textbf{G}\textbf{G}'}^2$ by the following equations
\begin{equation}
\tilde{\omega}_{\mathbf{\mathbf{G}\mathbf{G}'}}^2(\mathbf{q}) = \frac{\left | \lambda_{\mathbf{\mathbf{G}\mathbf{G}'}}
(\mathbf{q})  \right |}
{ \cos \phi_{\mathbf{GG}'} (\mathbf{q}) },
\end{equation}
\begin{equation}
\left| \lambda_{\mathbf{GG'}}(\mathbf{q}) \right | e^{i \phi_{\mathbf{GG}'}} (\mathbf{q}) =
\frac{\Omega_{\mathbf{GG}'}^{2}(\mathbf{q})}
{\delta_{\mathbf{GG'}} - \epsilon_{\mathbf{GG'}}^{-1} (\mathbf{q},0) },
\end{equation}
where $\lambda_{\mathbf{GG'}}(\mathbf{q})$ and $\phi_{\mathbf{GG}'} (\mathbf{q})$ are the amplitude and phase of the
renormalized
$\Omega_{\mathbf{GG'}}^{2}(\mathbf{q})$ respectively.
For computing the self-energy operator matrices it is necessary to construct the plane-wave matrix $M$ just like in the case
of the dielectric matrix.
For the terms $\Sigma_{SX}$ and $\Sigma_{CH}$ we have to choose the screened-Coulomb cut-off.
This cut-off has the same role of the dielectric cut-off, which is to define an energy truncation for the evaluation of the elements of $M$.
This energy has to be less or equal to  $E_{cut}$.
We chose it to be equal to  $E_{cut}$ in all calculations.
The convergence of the number of bands included in the summation of eq. \ref{BGW_CH_PPA}  has to be studied.

So we have to consider two convergence parameters: $E_{cut}$ (or number of bands $N_{\epsilon}$) of  $\epsilon_{\mathbf{GG'}}$
and the  number of bands $N_{\Sigma}$ included in $\Sigma_{CH}$.


\subsection{BSE calculations}

A detailed explanation of the BSE theory can be found in Ref. \onlinecite{PhysRevB.62.4927}.
For each excitonic state $S$, BSE can be written
\begin{equation}
\left( E_{c\mathbf{k}}^{QP} - E_{v\mathbf{k}}^{QP} \right) A_{vc\mathbf{k}}^S + \sum_{v'c'\mathbf{k}'}
\bra{vc\mathbf{k}} K_{eh} \ket{v'c'\mathbf{k}'} = \Omega^S A_{vc\mathbf{k}}^S,
\label{BGW_BSE}
\end{equation}
where $ E_{c\mathbf{k}}^{QP}$ and $ E_{v\mathbf{k}}^{QP}$ are the $G_0W_0$ QP energies for the  conduction ($c$) and valence bands ($v$) respectively,
$K_{eh}$ is the electron-hole kernel,
$A_{vc\mathbf{k}}^S$ and $\Omega^S$ the excitonic wavefunction and energy respectively for an  excitonic state $S$.
Tamm-Dancoff approximation\cite{PhysRev.78.382} is considered, where only valence to conduction band transitions are included.
The first step to solve the BSE is to compute the electron-hole kernel.
It can be separated in two terms $K_{eh} = K_d + K_x$, where $K_d$ is the screened direct interaction term and  $K_x$ is the bare exchange interaction term.
They are defined in the following way
\begin{multline}
\bra{vc\mathbf{k}} K_{d} \ket{v'c'\mathbf{k}'}  \\
= \sum_{\mathbf{GG}'} M_{c'c}^*(\mathbf{k},\mathbf{q},\mathbf{G})
W_{\mathbf{GG}'}(\mathbf{q},0) M_{v'v}(\mathbf{k},\mathbf{q},\mathbf{G}'),
\label{BGW_kernel_direct_GGspace}
\end{multline}
\begin{multline}
\bra{vc\mathbf{k}} K_{x} \ket{v'c'\mathbf{k}'} \\
=  \sum_{\mathbf{G} \neq 0 } M_{vc}^*(\mathbf{k},\mathbf{q},\mathbf{G})
v(\mathbf{q},0) M_{v'c'}(\mathbf{k},\mathbf{q},\mathbf{G})
\label{BGW_kernel_bare_GGspace}
\end{multline}
where the static screened Coulomb potential has the following expression
\begin{equation}
W_{\mathbf{GG'}}(\mathbf{q},0) = \epsilon^{-1}_{\mathbf{GG'}}(\mathbf{q},0)v(\mathbf{q}+\mathbf{G}')
\label{APBGW_static_W}.
\end{equation}
We also have to choose a cut-off for the kernel matrix construction because of matrices $M$.
The excitonic properties are very sensitive to the grid  of $\mathbf{k}$-points that is used because the contributions of $\mathbf{q} \to 0$ are very important.
We then have to compute the kernel in very fine grids, which is computationally prohibitively expensive.
The {\sc BerkeleyGW} package computes the kernel with a coarse grid of $\mathbf{k}$-points  and then interpolate it in a very fine grid of $\mathbf{k}$-points.
To perform this interpolation we have to choose a grid finer than the coarse grid and include a number of valence and conduction bands.
After solving the BSE,  the imaginary part of the dielectric function, which is proportional to absorption spectrum, can be
calculated by using the expression
\begin{equation}
\epsilon_2^{BSE}(\omega) = \frac{16 \pi^2 e^2}{\omega^2}\sum_S \left| \mathbf{e}  \cdot \bra{0}\mathbf{v}\ket{S}\right|^2 \delta(\omega - \Omega^S)
\label{epsilon_2_BGW}.
\end{equation}
The term  $\mathbf{e}  \cdot \bra{0}\mathbf{v}\ket{S}$ is  called the velocity matrix element in the direction of the polarization of the light $\mathbf{e}$.
We can also calculate the absorption spectrum without excitonic interactions with the expression
\begin{equation}
\epsilon_2^{RPA}(\omega) = \frac{16 \pi^2 e^2}{\omega^2} \sum_{vc\mathbf{k}}
\left| \mathbf{e}  \cdot \bra{v\mathbf{k}}\mathbf{v}\ket{c\mathbf{k}}\right|^2
\delta(\omega - E_{c\mathbf{k}}^{QP} + E_{v\mathbf{k}}^{QP})
\label{epsilon_2_BGW_neh1},
\end{equation}
which is a random phase approximation (RPA) calculation on top of the $G_0W_0$ results.

\subsection{Coulomb truncation}
\label{ssec:truncation}

To eliminate the artificial interactions between periodic images we can use the Coulomb truncation technique.
The Wigner-Seitz slab truncation implemented in {\sc BerkeleyGW} is used.
The Coulomb interaction is truncated at the edges of the unit cell in the direction perpendicular to the slab plane.
The bare Coulomb potential is then corrected to
\begin{equation}
v(\mathbf{q}) = \frac{4 \pi}{ q^2} \left(1-e^{-q_{xy}z_c} \cos\left(q_z z_c\right) \right)
\label{epsilon_2_BGW_neh},
\end{equation}
where $z_c$ is the truncation distance in the direction perpendicular to the phosphorene plane and
$q_{xy}$ and $q_z$ are the parallel and perpendicular components of the vector $\mathbf{q}$ respectively.

\section{Results and discussion}
\label{sec:results}
\subsection{DFT results}

\begin{figure}[t]
\centering
\includegraphics[scale=0.35]{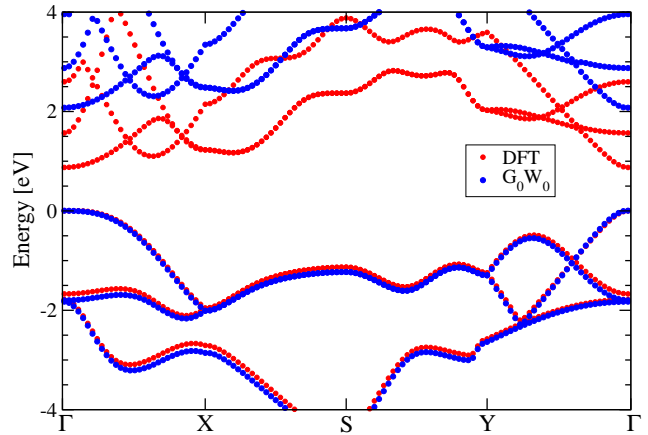}
\caption[Quasi-particle band-structure (blue) and DFT band-structure (red).]
{(Color online) QP band-structure (blue) and DFT band-structure (red). The QP bands were obtained with a $G_0W_0$ calculations
that used a $11 \times 11 \times 1$ $\mathbf{k}$-points grid and $L = 40$ bohr.}
\label{P_GW_band-struc}
\end{figure}

The DFT optimized lattice parameters for the phosphorene structure are $a=6.24$ bohr and $b=8.67$ bohr (see Fig.~\ref{P_structure}).
The electronic band structure is shown in Fig.~\ref{P_GW_band-struc}, and the band gap is direct at the $\Gamma$ point with an energy of 0.87 eV.
This value is in good agreement with several works that use the GGA as the exchange-correlation functional.\cite{PhysRevB.91.045433,highmobph,
PhysRevB.89.235319}
When HSE\cite{hseref} functionals are used, the band gap increases to about 1.5 eV.\cite{highmobph,PhysRevB.91.045433}
To our knowledge there are two experimental works where the band gap of phosphorene was measured, with
 values of 2.05 eV (Ref. \onlinecite{doi:10.1021/nl502892t}) and  $2.2 \pm 0.1$ eV (Ref. \onlinecite{natureHighly}).
As expected, these values are much larger than the DFT results, and confirm the need of using a
QP theory to have better prediction of the electronic properties of phosphorene.

\subsection{$G_0W_0$ results}

\label{sec:results_gw}

We calculated the QP band structure of phosphorene using the $G_0W_0$ approach.
Fig. \ref{P_GW_band-struc} shows the band structure of phosphorene with and without the QP corrections.
The QP band structure is essentially a shift of the DFT conduction bands and
the shape of the band structure changes little with the QP corrections.

First we studied the convergence of the QP band gap with the number of bands $N_{\Sigma}$ and $N_{\epsilon}$,
using a $9 \times 9 \times 1$ grid of $\mathbf{k}$-points and a vacuum size of 40 bohr.
Convergence was achieved for $N_{\epsilon}=1100$ bands for $\epsilon_{\mathbf{GG}'}$ (which corresponds to an $E_{cut}$ of 9.73 Ry)
and $N_{\Sigma}=1000$ bands for $\Sigma_{CH}$.
The band gap calculated is 2.18 eV, much larger than the DFT values and closer to the experimental.

\begin{figure}[t]
\centering
\includegraphics[scale=0.31]{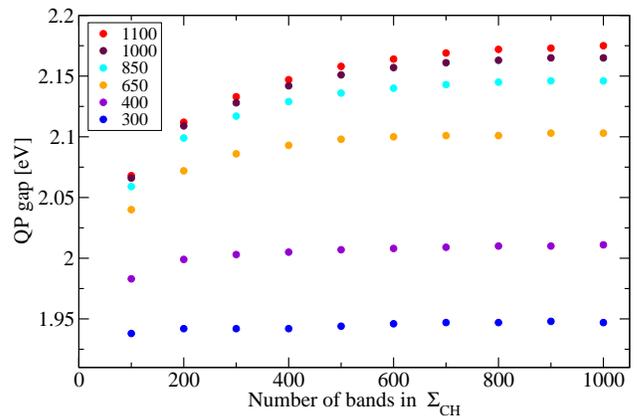}
\caption{(Color online) QP band gap vs $N_{\Sigma}$ (number of bands included in $\Sigma_{CH}$ summation) for a  $G_0W_0$ calculation
that used a grid of $9 \times 9 \times 1$  $\mathbf{k}$-points.
Different colors indicate the number of bands $N_{\epsilon}$ used to construct $\epsilon_{\mathbf{GG}'}$.
}
\label{p_gw_dap}
\end{figure}

Figure \ref{p_gw_dap} shows that the two convergence parameters are mutually dependent.
If we  choose to construct $\epsilon_{\mathbf{GG}'}$ with 300 bands, a convergence of 0.01~eV is achieved by including only about
400 bands in the $\Sigma_{CH}$ summation, and we would have a 1.95~eV band gap.
But if we choose $N_{\epsilon}=1100$ bands to construct the $\epsilon_{\mathbf{GG}'}$, we will need $N_{\Sigma}=1000$ bands
 to achieve the same convergence criterion, and would have a band gap of 2.18~eV.
These convergence parameters are interdependent  because few bands in the $\epsilon_{\mathbf{GG}'}$ calculation
 means including few $\mathbf{G}$ vectors, and so we are preventing the contribution from high energy empty bands
to the calculation of $\Sigma_{CH}$.\cite{PhysRevLett.105.146401}

\begin{figure}[t]
\centering
\includegraphics[scale=0.31]{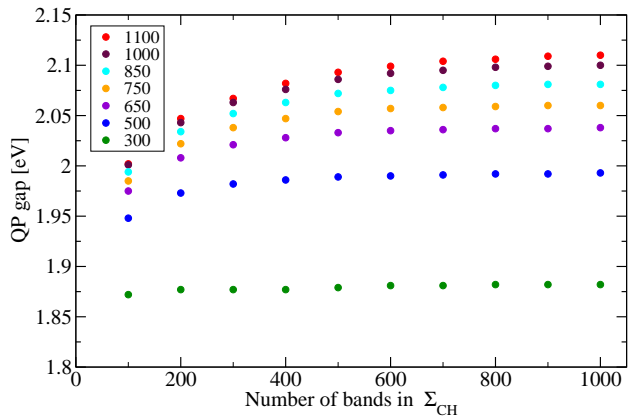}
\caption
{(Color online) QP band gap vs number of bands included in $\Sigma_{CH}$ summation for a  $G_0W_0$ calculation
that used a grid of $11 \times 11 \times 1$ $\mathbf{k}$-points.
Different colors indicate the number of bands used to construct $\epsilon_{\mathbf{GG}'}$.
 }
\label{p_gw_dap_11_11}
\end{figure}

$G_0W_0$ calculations also depend on the number of $\mathbf{k}$ or $\mathbf{q}$-points.
Despite convergence at DFT level being achieved for a small $\mathbf{k}$-points sampling,
for $G_0W_0$ it usually can only be achieved for a much denser grid.
This is due to the large variation of the dielectric function near the
$\Gamma$ point for bidimensional materials.\cite{PhysRevB.94.155406}
In order to capture this variation, a denser grid of $\mathbf{k}$-points is needed.
We repeated the previous calculations with a denser grid of $11 \times 11 \times 1$ $\mathbf{k}$-points.
Figure \ref{p_gw_dap_11_11} shows the convergence of the QP band gap with that grid.
Convergence was achieved for a QP band gap of 2.11 eV, which is only 0.07 eV different from the calculations with the smaller grid of
$9 \times 9 \times 1$ $\mathbf{k}$-points. This shows that for phosphorene $G_0W_0$ calculations are not so dependent on the density of $\mathbf{k}$-points.
To achieve a convergence of 0.01~eV we used even denser $\mathbf{k}$-point grids.
The results are shown in Table \ref{table_P_GW} alongside with results for different vacuum sizes.

%

\begin{table}[b]

\centering
\caption[Phosphorene $G_0W_0$ calculations  GPP obtained gap for all different vacuum sizes]
{$G_0W_0$ calculated gap [eV] for phosphorene for different vacuum sizes and $\mathbf{k}$-points sampling.}
\begin{ruledtabular}
\begin{tabular*}{\columnwidth}{@{\extracolsep{\fill} }lcccc}
\multirow{2}{*}{$\mathbf{k}$-points sampling} & \multicolumn{4}{c}{Vacuum size $L$ [bohr]}  \\

& 20 &30& 40 & 50  \\ \cline{2-5}\noalign{\smallskip}

$9 \times 9\times 1$ & 2.10 &2.12& 2.18 & 2.23   \\

$11 \times 11\times 1$ & 2.08 & 2.08 & 2.11 & 2.14   \\

$13 \times 13 \times 1$ & 2.07 & 2.06  & 2.07 & 2.10   \\

$15 \times 15 \times 1$ &  2.08 & 2.07 & 2.07 &  2.07 \\

$17 \times 17 \times 1$  &  2.08 & 2.07 & 2.06 & 2.06 \\

\end{tabular*}
\end{ruledtabular}
\label{table_P_GW}
\end{table}

It is well known that the QP band gap has a strong dependence on the vacuum size $L$ at the $G_0W_0$ level
 because of the non-local screening effects of the $GW$ approximation which makes the gap converge as $1/L$.\cite{INKSON197169}
A possible explanation for the increase of band gap with $L$ is that the screening becomes weaker
due to the electron-electron correlation being stronger.
The weaker the screening the larger the band gap.
In order to reduce this dependence, we use the truncation method mentioned in Subsection \ref{ssec:truncation}.
%
Table \ref{table_P_GW} shows that the QP band gap indeed increases with vacuum size, even with truncation.
We also made these calculations without truncation and we achieved the much lower value for the band gap of 1.81~eV and a stronger vacuum dependence.
This shows that using truncation method is indeed necessary to obtain the right results.
We also notice that when we increase the vacuum size, more $\mathbf{k}$-points are needed to achieve the convergence criterion of 0.01 eV.
The larger the vacuum size the larger the variation of the dielectric function near the
$\Gamma$ point, and so the denser the $\mathbf{k}$ point sampling needs to be.\cite{PhysRevB.88.245309}

In Table \ref{table:GW_summary_P} we summarize the results from other works.
The $G_0W_0$ QP band gap we obtained is in excellent agreement with the experimental ones.


\subsection{BSE results}
\label{ssec:BSE}
\begin{figure}[t]
\centering
\includegraphics[scale=0.35]{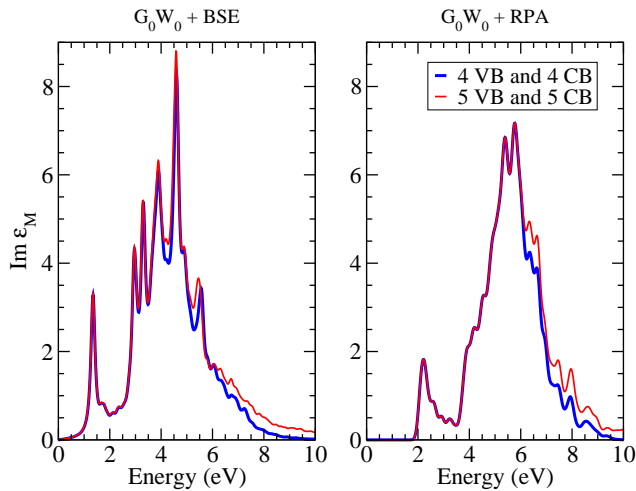}
\caption[Absorption spectrum of phosphorene with a fine grid of $24\times 24 \times 1$  $\mathbf{k}$-points with 4 and 5 VB.]
{(Color online)
Absorption spectrum of phosphorene with a fine grid of $24\times 24 \times 1$ $\mathbf{k}$-points.
Red curves: 4 valence bands and 4 conduction bands are included to construct the kernel matrix and for interpolation.
Blue curves: 5 valence bands and 5 conduction bands are included to construct the kernel matrix and for interpolation.
A gaussian broadening of 0.1 eV is used.}
\label{BSE_P_c30_24x24_3}
\end{figure}

The kernel matrix of the BSE calculations used a coarse grid of $13 \times 13 \times 1$ $\mathbf{k}$-points.
The starting point for our BSE calculations was chosen to be the $G_0W_0$ results with a vacuum size of 40 bohr.
For this calculation, we verified the convergence of the bands above and below the band gap, for the bands that were used for the absorption spectrum calculation.
We found that most of the energies were converged to better than 0.01 eV, with some cases with higher values, up to 0.014 eV, for the bands farther away from the gap.
We deemed this good enough, specially since the most relevant bands were well converged.
The first calculation was done with a fine grid for interpolation of $24 \times 24 \times 1$ $\mathbf{k}$-points.
We tested for 4 and 5 valence and conduction bands in the kernel matrix and for interpolation, and we found no difference
for energies below 4~eV, as can be seen in Fig. \ref{BSE_P_c30_24x24_3}.
So we only included 4 valence bands and 4 conduction bands for the construction of the kernel matrix and for interpolation.

\begin{figure}[t]
\centering
 \includegraphics[scale=0.35]{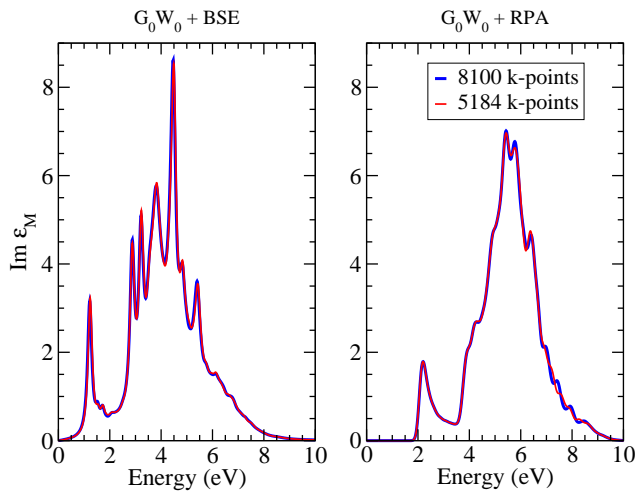}
\caption[Absorption spectra of phosphorene for different fine grids used on the interpolation (5184 $\mathbf{k}$-points and
8100 $\mathbf{k}$-points).]
{(Color online) Absorption spectra of phosphorene for different fine grids used on the interpolation.
Red curves: fine grid of 5184 $\mathbf{k}$-points
Blue curves: fine grid of 8100 $\mathbf{k}$-points.
We used 4 valence and 4 conduction bands for the construction of the kernel matrix and for the interpolation.
A gaussian broadening of 0.1 eV is used. For energies greater than 4 eV the spectra are not converged as discusses above.}
\label{BSE_P_c30_24x24_6}
\end{figure}

We also tested  the convergence with respect to the number of $\mathbf{k}$-points used for the interpolation.
Fig. \ref{BSE_P_c30_24x24_6} shows that both the BSE and the RPA absorption spectrum are converged for
a grid of $90\times 90 \times 1$ (8100) $\mathbf{k}$-points.

\begin{figure}[t]
\centering
\includegraphics[scale=0.3]{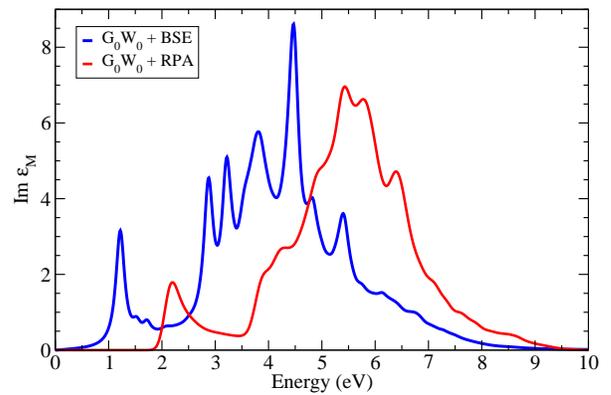}
\caption[Absorption spectrum of phosphorene with  a fine grid of $90 \times 90 \times 1$ $\mathbf{k}$-points.]
{(Color online) Absorption spectrum of phosphorene.
This calculation is computed in a coarse grid of $13 \times 13 \times 1$  $\mathbf{k}$-points and then
interpolated in a finer grid of $90 \times 90 \times 1$ $\mathbf{k}$-points.
We used 4 valence and 4 conduction bands for the construction of the kernel matrix and for the interpolation.  A gaussian broadening of
0.1 eV is used.}
\label{BSE_P_c40_96x96}
\end{figure}

In Fig. \ref{BSE_P_c40_96x96} a comparison between the BSE and the RPA spectra is shown.
When excitonic effects are included, the spectrum red-shifts and discrete peaks of the excitonic states appear, as expected.
The first excitonic peak is at an energy of 1.22 eV.
This is the optical gap of phosphorene.
Knowing that the optical gap is 1.22 eV and the $G_0W_0$ QP band gap is 2.06 eV, the excitonic binding energy is 0.84 eV.
This value is in agreement with the experimental one of $0.9 \pm 0.12$ eV (Ref. \onlinecite{natureHighly}).
A high excitonic binding energy is expected in this material because of its low dimensionality and low screening.
The low dimensionality increases the confinement between the electron and hole which enhances its Coulomb interaction.

Table \ref{table:BSE_P_summary} summarizes the results of our work and three other works that we found that use $G_0W_0+$BSE approach.
It also shows the only experimental value we could find.
The optical gap we obtained is a bit smaller than the experimental value.
This may be due to substrate effects in the experiment; our calculation assumes a suspended monolayer while the experimental measurements were done on a substrate.
Our results also differ a bit from other theoretical works.
It is not clear to us if  $G_0W_0$ calculations were fully converged in other works as we discussed above.
And our results showed that a grid of $90 \times 90 \times 1$ $\mathbf{k}$-points is needed in order
to achieve convergence, much larger than the ones used by others.

\section{Conclusion}
\label{sec:conclusion}

$GW$ approximation and the BSE methods were used to study the electronic and optical properties of phosphorene in a careful and systematic numerical convergence study.
We found that the $G_0W_0$ QP band gap is 2.06 eV, closer to the experimental values than the results from other works.
We also did BSE calculations in order to include the excitonic effects which are important in materials with low dimensionality.
An optical gap of 1.22 eV was obtained and an excitonic binding energy of 0.84 eV.
These values are consistent with a recent experimental work and results from other theoretical works.
We also saw that in order to achieve convergence using those techniques we have to consider the interdependence of the parameters
that have to be converged.

\section*{Acknowledgments}

R.M. Ribeiro acknowledge support from the European Commission through
the project ``Graphene-Driven Revolutions in ICT and Beyond" (Ref. No. 696656),
COMPETE2020, PORTUGAL2020, FEDER and the Portuguese Foundation for Science and Technology
(FCT) through project PTDC/FIS-NAN/3668/2014 and in the framework of the Strategic Financing UID/FIS/04650/2013.


\begin{thebibliography}{36}%
\makeatletter
\providecommand \@ifxundefined [1]{%
 \@ifx{#1\undefined}
}%
\providecommand \@ifnum [1]{%
 \ifnum #1\expandafter \@firstoftwo
 \else \expandafter \@secondoftwo
 \fi
}%
\providecommand \@ifx [1]{%
 \ifx #1\expandafter \@firstoftwo
 \else \expandafter \@secondoftwo
 \fi
}%
\providecommand \natexlab [1]{#1}%
\providecommand \enquote  [1]{``#1''}%
\providecommand \bibnamefont  [1]{#1}%
\providecommand \bibfnamefont [1]{#1}%
\providecommand \citenamefont [1]{#1}%
\providecommand \href@noop [0]{\@secondoftwo}%
\providecommand \href [0]{\begingroup \@sanitize@url \@href}%
\providecommand \@href[1]{\@@startlink{#1}\@@href}%
\providecommand \@@href[1]{\endgroup#1\@@endlink}%
\providecommand \@sanitize@url [0]{\catcode `\\12\catcode `\$12\catcode
  `\&12\catcode `\#12\catcode `\^12\catcode `\_12\catcode `\%12\relax}%
\providecommand \@@startlink[1]{}%
\providecommand \@@endlink[0]{}%
\providecommand \url  [0]{\begingroup\@sanitize@url \@url }%
\providecommand \@url [1]{\endgroup\@href {#1}{\urlprefix }}%
\providecommand \urlprefix  [0]{URL }%
\providecommand \Eprint [0]{\href }%
\providecommand \doibase [0]{http://dx.doi.org/}%
\providecommand \selectlanguage [0]{\@gobble}%
\providecommand \bibinfo  [0]{\@secondoftwo}%
\providecommand \bibfield  [0]{\@secondoftwo}%
\providecommand \translation [1]{[#1]}%
\providecommand \BibitemOpen [0]{}%
\providecommand \bibitemStop [0]{}%
\providecommand \bibitemNoStop [0]{.\EOS\space}%
\providecommand \EOS [0]{\spacefactor3000\relax}%
\providecommand \BibitemShut  [1]{\csname bibitem#1\endcsname}%
\let\auto@bib@innerbib\@empty
\bibitem [{\citenamefont {Liang}\ \emph {et~al.}(2014)\citenamefont {Liang},
  \citenamefont {Wang}, \citenamefont {Lin}, \citenamefont {Sumpter},
  \citenamefont {Meunier},\ and\ \citenamefont {Pan}}]{doi:10.1021/nl502892t}%
  \BibitemOpen
  \bibfield  {author} {\bibinfo {author} {\bibfnamefont {L.}~\bibnamefont
  {Liang}}, \bibinfo {author} {\bibfnamefont {J.}~\bibnamefont {Wang}},
  \bibinfo {author} {\bibfnamefont {W.}~\bibnamefont {Lin}}, \bibinfo {author}
  {\bibfnamefont {B.~G.}\ \bibnamefont {Sumpter}}, \bibinfo {author}
  {\bibfnamefont {V.}~\bibnamefont {Meunier}}, \ and\ \bibinfo {author}
  {\bibfnamefont {M.}~\bibnamefont {Pan}},\ }\href {\doibase 10.1021/nl502892t}
  {\bibfield  {journal} {\bibinfo  {journal} {Nano Letters}\ }\textbf {\bibinfo
  {volume} {14}},\ \bibinfo {pages} {6400} (\bibinfo {year} {2014})},\ \bibinfo
  {note} {pMID: 25343376}\BibitemShut {NoStop}%
\bibitem [{\citenamefont {Keyes}(1953)}]{PhysRev.92.580}%
  \BibitemOpen
  \bibfield  {author} {\bibinfo {author} {\bibfnamefont {R.~W.}\ \bibnamefont
  {Keyes}},\ }\href {\doibase 10.1103/PhysRev.92.580} {\bibfield  {journal}
  {\bibinfo  {journal} {Phys. Rev.}\ }\textbf {\bibinfo {volume} {92}},\
  \bibinfo {pages} {580} (\bibinfo {year} {1953})}\BibitemShut {NoStop}%
\bibitem [{\citenamefont {Qiao}\ \emph {et~al.}(2014)\citenamefont {Qiao},
  \citenamefont {Kong}, \citenamefont {Hu}, \citenamefont {Yang},\ and\
  \citenamefont {Ji}}]{highmobph}%
  \BibitemOpen
  \bibfield  {author} {\bibinfo {author} {\bibfnamefont {J.}~\bibnamefont
  {Qiao}}, \bibinfo {author} {\bibfnamefont {X.}~\bibnamefont {Kong}}, \bibinfo
  {author} {\bibfnamefont {Z.-X.}\ \bibnamefont {Hu}}, \bibinfo {author}
  {\bibfnamefont {F.}~\bibnamefont {Yang}}, \ and\ \bibinfo {author}
  {\bibfnamefont {W.}~\bibnamefont {Ji}},\ }\href {\doibase 10.1038/ncomms5475}
  {\bibfield  {journal} {\bibinfo  {journal} {Nat Commun}\ }\textbf {\bibinfo
  {volume} {5}} (\bibinfo {year} {2014}),\ 10.1038/ncomms5475}\BibitemShut
  {NoStop}%
\bibitem [{\citenamefont {Liu}\ \emph {et~al.}(2014)\citenamefont {Liu},
  \citenamefont {Neal}, \citenamefont {Zhu}, \citenamefont {Luo}, \citenamefont
  {Xu}, \citenamefont {Tománek},\ and\ \citenamefont
  {Ye}}]{doi:10.1021/nn501226z}%
  \BibitemOpen
  \bibfield  {author} {\bibinfo {author} {\bibfnamefont {H.}~\bibnamefont
  {Liu}}, \bibinfo {author} {\bibfnamefont {A.~T.}\ \bibnamefont {Neal}},
  \bibinfo {author} {\bibfnamefont {Z.}~\bibnamefont {Zhu}}, \bibinfo {author}
  {\bibfnamefont {Z.}~\bibnamefont {Luo}}, \bibinfo {author} {\bibfnamefont
  {X.}~\bibnamefont {Xu}}, \bibinfo {author} {\bibfnamefont {D.}~\bibnamefont
  {Tománek}}, \ and\ \bibinfo {author} {\bibfnamefont {P.~D.}\ \bibnamefont
  {Ye}},\ }\href {\doibase 10.1021/nn501226z} {\bibfield  {journal} {\bibinfo
  {journal} {ACS Nano}\ }\textbf {\bibinfo {volume} {8}},\ \bibinfo {pages}
  {4033} (\bibinfo {year} {2014})},\ \bibinfo {note} {pMID:
  24655084}\BibitemShut {NoStop}%
\bibitem [{\citenamefont {Li}\ \emph {et~al.}(2014)\citenamefont {Li},
  \citenamefont {Yu}, \citenamefont {Ye}, \citenamefont {Ge}, \citenamefont
  {Ou}, \citenamefont {Wu}, \citenamefont {Feng}, \citenamefont {Chen},\ and\
  \citenamefont {Zhang}}]{Li2014}%
  \BibitemOpen
  \bibfield  {author} {\bibinfo {author} {\bibfnamefont {L.}~\bibnamefont
  {Li}}, \bibinfo {author} {\bibfnamefont {Y.}~\bibnamefont {Yu}}, \bibinfo
  {author} {\bibfnamefont {G.~J.}\ \bibnamefont {Ye}}, \bibinfo {author}
  {\bibfnamefont {Q.}~\bibnamefont {Ge}}, \bibinfo {author} {\bibfnamefont
  {X.}~\bibnamefont {Ou}}, \bibinfo {author} {\bibfnamefont {H.}~\bibnamefont
  {Wu}}, \bibinfo {author} {\bibfnamefont {D.}~\bibnamefont {Feng}}, \bibinfo
  {author} {\bibfnamefont {X.~H.}\ \bibnamefont {Chen}}, \ and\ \bibinfo
  {author} {\bibfnamefont {Y.}~\bibnamefont {Zhang}},\ }\href
  {http://dx.doi.org/10.1038/nnano.2014.35} {\bibfield  {journal} {\bibinfo
  {journal} {Nat Nano}\ }\textbf {\bibinfo {volume} {9}},\ \bibinfo {pages}
  {372} (\bibinfo {year} {2014})},\ \bibinfo {note} {article}\BibitemShut
  {NoStop}%
\bibitem [{\citenamefont {Tran}\ \emph {et~al.}(2014)\citenamefont {Tran},
  \citenamefont {Soklaski}, \citenamefont {Liang},\ and\ \citenamefont
  {Yang}}]{PhysRevB.89.235319}%
  \BibitemOpen
  \bibfield  {author} {\bibinfo {author} {\bibfnamefont {V.}~\bibnamefont
  {Tran}}, \bibinfo {author} {\bibfnamefont {R.}~\bibnamefont {Soklaski}},
  \bibinfo {author} {\bibfnamefont {Y.}~\bibnamefont {Liang}}, \ and\ \bibinfo
  {author} {\bibfnamefont {L.}~\bibnamefont {Yang}},\ }\href {\doibase
  10.1103/PhysRevB.89.235319} {\bibfield  {journal} {\bibinfo  {journal} {Phys.
  Rev. B}\ }\textbf {\bibinfo {volume} {89}},\ \bibinfo {pages} {235319}
  (\bibinfo {year} {2014})}\BibitemShut {NoStop}%
\bibitem [{\citenamefont {Zhang}\ \emph {et~al.}(2017)\citenamefont {Zhang},
  \citenamefont {Huang}, \citenamefont {Chaves}, \citenamefont {Özçelik},
  \citenamefont {Low},\ and\ \citenamefont {Yan}}]{ncomms14071}%
  \BibitemOpen
  \bibfield  {author} {\bibinfo {author} {\bibfnamefont {G.}~\bibnamefont
  {Zhang}}, \bibinfo {author} {\bibfnamefont {S.}~\bibnamefont {Huang}},
  \bibinfo {author} {\bibfnamefont {C.}~\bibnamefont {Chaves}, \bibfnamefont
  {A.and~Song}}, \bibinfo {author} {\bibfnamefont {V.}~\bibnamefont
  {Özçelik}}, \bibinfo {author} {\bibfnamefont {T.}~\bibnamefont {Low}}, \
  and\ \bibinfo {author} {\bibfnamefont {H.}~\bibnamefont {Yan}},\ }\href
  {http://dx.doi.org/10.1038/ncomms14071} {\bibfield  {journal} {\bibinfo
  {journal} {Nature Communications}\ }\textbf {\bibinfo {volume} {8}},\
  \bibinfo {pages} {14071} (\bibinfo {year} {2017})}\BibitemShut {NoStop}%
\bibitem [{\citenamefont {\"Oz\ifmmode~\mbox{\c{c}}\else \c{c}\fi{}elik}\ \emph
  {et~al.}(2016)\citenamefont {\"Oz\ifmmode~\mbox{\c{c}}\else \c{c}\fi{}elik},
  \citenamefont {Azadani}, \citenamefont {Yang}, \citenamefont {Koester},\ and\
  \citenamefont {Low}}]{PhysRevB.94.035125}%
  \BibitemOpen
  \bibfield  {author} {\bibinfo {author} {\bibfnamefont {V.~O.}\ \bibnamefont
  {\"Oz\ifmmode~\mbox{\c{c}}\else \c{c}\fi{}elik}}, \bibinfo {author}
  {\bibfnamefont {J.~G.}\ \bibnamefont {Azadani}}, \bibinfo {author}
  {\bibfnamefont {C.}~\bibnamefont {Yang}}, \bibinfo {author} {\bibfnamefont
  {S.~J.}\ \bibnamefont {Koester}}, \ and\ \bibinfo {author} {\bibfnamefont
  {T.}~\bibnamefont {Low}},\ }\href {\doibase 10.1103/PhysRevB.94.035125}
  {\bibfield  {journal} {\bibinfo  {journal} {Phys. Rev. B}\ }\textbf {\bibinfo
  {volume} {94}},\ \bibinfo {pages} {035125} (\bibinfo {year}
  {2016})}\BibitemShut {NoStop}%
\bibitem [{\citenamefont {Ganesan}\ \emph {et~al.}(2016)\citenamefont
  {Ganesan}, \citenamefont {Linghu}, \citenamefont {Zhang},\ and\ \citenamefont
  {Feng}}]{APS2016}%
  \BibitemOpen
  \bibfield  {author} {\bibinfo {author} {\bibfnamefont {V.~D.~S.}\
  \bibnamefont {Ganesan}}, \bibinfo {author} {\bibfnamefont {J.}~\bibnamefont
  {Linghu}}, \bibinfo {author} {\bibfnamefont {C.}~\bibnamefont {Zhang}}, \
  and\ \bibinfo {author} {\bibfnamefont {Y.~P.}\ \bibnamefont {Feng}},\ }\href
  {\doibase 10.1063/1.4944642} {\bibfield  {journal} {\bibinfo  {journal}
  {Applied Physics Letters}\ }\textbf {\bibinfo {volume} {108}},\ \bibinfo
  {pages} {122105} (\bibinfo {year} {2016})}\BibitemShut {NoStop}%
\bibitem [{\citenamefont {\ifmmode \mbox{\c{C}}\else \c{C}\fi{}ak\ifmmode
  \imath \else~\i \fi{}r}\ \emph {et~al.}(2014)\citenamefont {\ifmmode
  \mbox{\c{C}}\else \c{C}\fi{}ak\ifmmode \imath \else~\i \fi{}r}, \citenamefont
  {Sahin},\ and\ \citenamefont {Peeters}}]{PhysRevB.90.205421}%
  \BibitemOpen
  \bibfield  {author} {\bibinfo {author} {\bibfnamefont {D.}~\bibnamefont
  {\ifmmode \mbox{\c{C}}\else \c{C}\fi{}ak\ifmmode \imath \else~\i \fi{}r}},
  \bibinfo {author} {\bibfnamefont {H.}~\bibnamefont {Sahin}}, \ and\ \bibinfo
  {author} {\bibfnamefont {F.~M.}\ \bibnamefont {Peeters}},\ }\href {\doibase
  10.1103/PhysRevB.90.205421} {\bibfield  {journal} {\bibinfo  {journal} {Phys.
  Rev. B}\ }\textbf {\bibinfo {volume} {90}},\ \bibinfo {pages} {205421}
  (\bibinfo {year} {2014})}\BibitemShut {NoStop}%
\bibitem [{\citenamefont {Fei}\ and\ \citenamefont
  {Yang}(2014)}]{doi:10.1021/nl500935z}%
  \BibitemOpen
  \bibfield  {author} {\bibinfo {author} {\bibfnamefont {R.}~\bibnamefont
  {Fei}}\ and\ \bibinfo {author} {\bibfnamefont {L.}~\bibnamefont {Yang}},\
  }\href {\doibase 10.1021/nl500935z} {\bibfield  {journal} {\bibinfo
  {journal} {Nano Letters}\ }\textbf {\bibinfo {volume} {14}},\ \bibinfo
  {pages} {2884} (\bibinfo {year} {2014})},\ \bibinfo {note} {pMID:
  24779386}\BibitemShut {NoStop}%
\bibitem [{\citenamefont {Jiang}\ and\ \citenamefont {Park}(2014)}]{Jiang2014}%
  \BibitemOpen
  \bibfield  {author} {\bibinfo {author} {\bibfnamefont {J.-W.}\ \bibnamefont
  {Jiang}}\ and\ \bibinfo {author} {\bibfnamefont {H.~S.}\ \bibnamefont
  {Park}},\ }\href {http://dx.doi.org/10.1038/ncomms5727} {\bibfield  {journal}
  {\bibinfo  {journal} {Nature Communications}\ }\textbf {\bibinfo {volume}
  {5}},\ \bibinfo {pages} {4727 EP } (\bibinfo {year} {2014})},\ \bibinfo
  {note} {article}\BibitemShut {NoStop}%
\bibitem [{\citenamefont {Wei}\ and\ \citenamefont
  {Peng}(2014)}]{doi:10.1063/1.4885215}%
  \BibitemOpen
  \bibfield  {author} {\bibinfo {author} {\bibfnamefont {Q.}~\bibnamefont
  {Wei}}\ and\ \bibinfo {author} {\bibfnamefont {X.}~\bibnamefont {Peng}},\
  }\href {\doibase 10.1063/1.4885215} {\bibfield  {journal} {\bibinfo
  {journal} {Applied Physics Letters}\ }\textbf {\bibinfo {volume} {104}},\
  \bibinfo {pages} {251915} (\bibinfo {year} {2014})}\BibitemShut {NoStop}%
\bibitem [{\citenamefont {Kou}\ \emph {et~al.}(2014)\citenamefont {Kou},
  \citenamefont {Frauenheim},\ and\ \citenamefont
  {Chen}}]{doi:10.1021/jz501188k}%
  \BibitemOpen
  \bibfield  {author} {\bibinfo {author} {\bibfnamefont {L.}~\bibnamefont
  {Kou}}, \bibinfo {author} {\bibfnamefont {T.}~\bibnamefont {Frauenheim}}, \
  and\ \bibinfo {author} {\bibfnamefont {C.}~\bibnamefont {Chen}},\ }\href
  {\doibase 10.1021/jz501188k} {\bibfield  {journal} {\bibinfo  {journal} {The
  Journal of Physical Chemistry Letters}\ }\textbf {\bibinfo {volume} {5}},\
  \bibinfo {pages} {2675} (\bibinfo {year} {2014})},\ \bibinfo {note} {pMID:
  26277962}\BibitemShut {NoStop}%
\bibitem [{\citenamefont {Deng}\ \emph {et~al.}(2014)\citenamefont {Deng},
  \citenamefont {Luo}, \citenamefont {Conrad}, \citenamefont {Liu},
  \citenamefont {Gong}, \citenamefont {Najmaei}, \citenamefont {Ajayan},
  \citenamefont {Lou}, \citenamefont {Xu},\ and\ \citenamefont
  {Ye}}]{doi:10.1021/nn5027388}%
  \BibitemOpen
  \bibfield  {author} {\bibinfo {author} {\bibfnamefont {Y.}~\bibnamefont
  {Deng}}, \bibinfo {author} {\bibfnamefont {Z.}~\bibnamefont {Luo}}, \bibinfo
  {author} {\bibfnamefont {N.~J.}\ \bibnamefont {Conrad}}, \bibinfo {author}
  {\bibfnamefont {H.}~\bibnamefont {Liu}}, \bibinfo {author} {\bibfnamefont
  {Y.}~\bibnamefont {Gong}}, \bibinfo {author} {\bibfnamefont {S.}~\bibnamefont
  {Najmaei}}, \bibinfo {author} {\bibfnamefont {P.~M.}\ \bibnamefont {Ajayan}},
  \bibinfo {author} {\bibfnamefont {J.}~\bibnamefont {Lou}}, \bibinfo {author}
  {\bibfnamefont {X.}~\bibnamefont {Xu}}, \ and\ \bibinfo {author}
  {\bibfnamefont {P.~D.}\ \bibnamefont {Ye}},\ }\href {\doibase
  10.1021/nn5027388} {\bibfield  {journal} {\bibinfo  {journal} {ACS Nano}\
  }\textbf {\bibinfo {volume} {8}},\ \bibinfo {pages} {8292} (\bibinfo {year}
  {2014})},\ \bibinfo {note} {pMID: 25019534}\BibitemShut {NoStop}%
\bibitem [{\citenamefont {Hedin}(1965)}]{PhysRev.139.A796}%
  \BibitemOpen
  \bibfield  {author} {\bibinfo {author} {\bibfnamefont {L.}~\bibnamefont
  {Hedin}},\ }\href {\doibase 10.1103/PhysRev.139.A796} {\bibfield  {journal}
  {\bibinfo  {journal} {Phys. Rev.}\ }\textbf {\bibinfo {volume} {139}},\
  \bibinfo {pages} {A796} (\bibinfo {year} {1965})}\BibitemShut {NoStop}%
\bibitem [{\citenamefont {Hedin}\ and\ \citenamefont {S.}(1969)}]{HedinS}%
  \BibitemOpen
  \bibfield  {author} {\bibinfo {author} {\bibfnamefont {L.}~\bibnamefont
  {Hedin}}\ and\ \bibinfo {author} {\bibfnamefont {L.}~\bibnamefont {S.}},\
  }\href@noop {} {\bibfield  {journal} {\bibinfo  {journal} {Solid State
  Physics}\ }\textbf {\bibinfo {volume} {23}} (\bibinfo {year}
  {1969})}\BibitemShut {NoStop}%
\bibitem [{\citenamefont {Salpeter}\ and\ \citenamefont
  {Bethe}(1951)}]{PhysRev.84.1232}%
  \BibitemOpen
  \bibfield  {author} {\bibinfo {author} {\bibfnamefont {E.~E.}\ \bibnamefont
  {Salpeter}}\ and\ \bibinfo {author} {\bibfnamefont {H.~A.}\ \bibnamefont
  {Bethe}},\ }\href {\doibase 10.1103/PhysRev.84.1232} {\bibfield  {journal}
  {\bibinfo  {journal} {Phys. Rev.}\ }\textbf {\bibinfo {volume} {84}},\
  \bibinfo {pages} {1232} (\bibinfo {year} {1951})}\BibitemShut {NoStop}%
\bibitem [{\citenamefont {Albrecht}\ \emph {et~al.}(1998)\citenamefont
  {Albrecht}, \citenamefont {Reining}, \citenamefont {Del~Sole},\ and\
  \citenamefont {Onida}}]{PhysRevLett.80.4510}%
  \BibitemOpen
  \bibfield  {author} {\bibinfo {author} {\bibfnamefont {S.}~\bibnamefont
  {Albrecht}}, \bibinfo {author} {\bibfnamefont {L.}~\bibnamefont {Reining}},
  \bibinfo {author} {\bibfnamefont {R.}~\bibnamefont {Del~Sole}}, \ and\
  \bibinfo {author} {\bibfnamefont {G.}~\bibnamefont {Onida}},\ }\href
  {\doibase 10.1103/PhysRevLett.80.4510} {\bibfield  {journal} {\bibinfo
  {journal} {Phys. Rev. Lett.}\ }\textbf {\bibinfo {volume} {80}},\ \bibinfo
  {pages} {4510} (\bibinfo {year} {1998})}\BibitemShut {NoStop}%
\bibitem [{\citenamefont {Rudenko}\ \emph {et~al.}(2015)\citenamefont
  {Rudenko}, \citenamefont {Yuan},\ and\ \citenamefont
  {Katsnelson}}]{PhysRevB.92.085419}%
  \BibitemOpen
  \bibfield  {author} {\bibinfo {author} {\bibfnamefont {A.~N.}\ \bibnamefont
  {Rudenko}}, \bibinfo {author} {\bibfnamefont {S.}~\bibnamefont {Yuan}}, \
  and\ \bibinfo {author} {\bibfnamefont {M.~I.}\ \bibnamefont {Katsnelson}},\
  }\href {\doibase 10.1103/PhysRevB.92.085419} {\bibfield  {journal} {\bibinfo
  {journal} {Phys. Rev. B}\ }\textbf {\bibinfo {volume} {92}},\ \bibinfo
  {pages} {085419} (\bibinfo {year} {2015})}\BibitemShut {NoStop}%
\bibitem [{\citenamefont {Rudenko}\ and\ \citenamefont
  {Katsnelson}(2014)}]{PhysRevB.89.201408}%
  \BibitemOpen
  \bibfield  {author} {\bibinfo {author} {\bibfnamefont {A.~N.}\ \bibnamefont
  {Rudenko}}\ and\ \bibinfo {author} {\bibfnamefont {M.~I.}\ \bibnamefont
  {Katsnelson}},\ }\href {\doibase 10.1103/PhysRevB.89.201408} {\bibfield
  {journal} {\bibinfo  {journal} {Phys. Rev. B}\ }\textbf {\bibinfo {volume}
  {89}},\ \bibinfo {pages} {201408} (\bibinfo {year} {2014})}\BibitemShut
  {NoStop}%
\bibitem [{\citenamefont {Tran}\ \emph {et~al.}(2015)\citenamefont {Tran},
  \citenamefont {Fei},\ and\ \citenamefont {Yang}}]{2053-1583-2-4-044014}%
  \BibitemOpen
  \bibfield  {author} {\bibinfo {author} {\bibfnamefont {V.}~\bibnamefont
  {Tran}}, \bibinfo {author} {\bibfnamefont {R.}~\bibnamefont {Fei}}, \ and\
  \bibinfo {author} {\bibfnamefont {L.}~\bibnamefont {Yang}},\ }\href
  {http://stacks.iop.org/2053-1583/2/i=4/a=044014} {\bibfield  {journal}
  {\bibinfo  {journal} {2D Materials}\ }\textbf {\bibinfo {volume} {2}},\
  \bibinfo {pages} {044014} (\bibinfo {year} {2015})}\BibitemShut {NoStop}%
\bibitem [{\citenamefont {Wang}\ \emph
  {et~al.}(2015{\natexlab{a}})\citenamefont {Wang}, \citenamefont {Jones},
  \citenamefont {Seyler}, \citenamefont {Tran}, \citenamefont {Jia},
  \citenamefont {Zhao}, \citenamefont {Wang}, \citenamefont {Yang},
  \citenamefont {Xu},\ and\ \citenamefont {Xia}}]{natureHighly}%
  \BibitemOpen
  \bibfield  {author} {\bibinfo {author} {\bibfnamefont {X.}~\bibnamefont
  {Wang}}, \bibinfo {author} {\bibfnamefont {A.~M.}\ \bibnamefont {Jones}},
  \bibinfo {author} {\bibfnamefont {K.~L.}\ \bibnamefont {Seyler}}, \bibinfo
  {author} {\bibfnamefont {V.}~\bibnamefont {Tran}}, \bibinfo {author}
  {\bibfnamefont {Y.}~\bibnamefont {Jia}}, \bibinfo {author} {\bibfnamefont
  {H.}~\bibnamefont {Zhao}}, \bibinfo {author} {\bibfnamefont {H.}~\bibnamefont
  {Wang}}, \bibinfo {author} {\bibfnamefont {L.}~\bibnamefont {Yang}}, \bibinfo
  {author} {\bibfnamefont {X.}~\bibnamefont {Xu}}, \ and\ \bibinfo {author}
  {\bibfnamefont {F.}~\bibnamefont {Xia}},\ }\href {\doibase
  10.1038/nnano.2015.71} {\bibfield  {journal} {\bibinfo  {journal} {Nat Nano}\
  }\textbf {\bibinfo {volume} {10}},\ \bibinfo {pages} {517} (\bibinfo {year}
  {2015}{\natexlab{a}})}\BibitemShut {NoStop}%
\bibitem [{\citenamefont {Shih}\ \emph {et~al.}(2010)\citenamefont {Shih},
  \citenamefont {Xue}, \citenamefont {Zhang}, \citenamefont {Cohen},\ and\
  \citenamefont {Louie}}]{PhysRevLett.105.146401}%
  \BibitemOpen
  \bibfield  {author} {\bibinfo {author} {\bibfnamefont {B.-C.}\ \bibnamefont
  {Shih}}, \bibinfo {author} {\bibfnamefont {Y.}~\bibnamefont {Xue}}, \bibinfo
  {author} {\bibfnamefont {P.}~\bibnamefont {Zhang}}, \bibinfo {author}
  {\bibfnamefont {M.~L.}\ \bibnamefont {Cohen}}, \ and\ \bibinfo {author}
  {\bibfnamefont {S.~G.}\ \bibnamefont {Louie}},\ }\href {\doibase
  10.1103/PhysRevLett.105.146401} {\bibfield  {journal} {\bibinfo  {journal}
  {Phys. Rev. Lett.}\ }\textbf {\bibinfo {volume} {105}},\ \bibinfo {pages}
  {146401} (\bibinfo {year} {2010})}\BibitemShut {NoStop}%
\bibitem [{\citenamefont {Giannozzi}\ \emph {et~al.}(2009)\citenamefont
  {Giannozzi}, \citenamefont {Baroni}, \citenamefont {Bonini}, \citenamefont
  {Calandra}, \citenamefont {Car}, \citenamefont {Cavazzoni}, \citenamefont
  {Ceresoli}, \citenamefont {Chiarotti}, \citenamefont {Cococcioni},
  \citenamefont {Dabo}, \citenamefont {{Dal Corso}}, \citenamefont
  {de~Gironcoli}, \citenamefont {Fabris}, \citenamefont {Fratesi},
  \citenamefont {Gebauer}, \citenamefont {Gerstmann}, \citenamefont
  {Gougoussis}, \citenamefont {Kokalj}, \citenamefont {Lazzeri}, \citenamefont
  {Martin-Samos}, \citenamefont {Marzari}, \citenamefont {Mauri}, \citenamefont
  {Mazzarello}, \citenamefont {Paolini}, \citenamefont {Pasquarello},
  \citenamefont {Paulatto}, \citenamefont {Sbraccia}, \citenamefont {Scandolo},
  \citenamefont {Sclauzero}, \citenamefont {Seitsonen}, \citenamefont
  {Smogunov}, \citenamefont {Umari},\ and\ \citenamefont
  {Wentzcovitch}}]{QE-2009}%
  \BibitemOpen
  \bibfield  {author} {\bibinfo {author} {\bibfnamefont {P.}~\bibnamefont
  {Giannozzi}}, \bibinfo {author} {\bibfnamefont {S.}~\bibnamefont {Baroni}},
  \bibinfo {author} {\bibfnamefont {N.}~\bibnamefont {Bonini}}, \bibinfo
  {author} {\bibfnamefont {M.}~\bibnamefont {Calandra}}, \bibinfo {author}
  {\bibfnamefont {R.}~\bibnamefont {Car}}, \bibinfo {author} {\bibfnamefont
  {C.}~\bibnamefont {Cavazzoni}}, \bibinfo {author} {\bibfnamefont
  {D.}~\bibnamefont {Ceresoli}}, \bibinfo {author} {\bibfnamefont {G.~L.}\
  \bibnamefont {Chiarotti}}, \bibinfo {author} {\bibfnamefont {M.}~\bibnamefont
  {Cococcioni}}, \bibinfo {author} {\bibfnamefont {I.}~\bibnamefont {Dabo}},
  \bibinfo {author} {\bibfnamefont {A.}~\bibnamefont {{Dal Corso}}}, \bibinfo
  {author} {\bibfnamefont {S.}~\bibnamefont {de~Gironcoli}}, \bibinfo {author}
  {\bibfnamefont {S.}~\bibnamefont {Fabris}}, \bibinfo {author} {\bibfnamefont
  {G.}~\bibnamefont {Fratesi}}, \bibinfo {author} {\bibfnamefont
  {R.}~\bibnamefont {Gebauer}}, \bibinfo {author} {\bibfnamefont
  {U.}~\bibnamefont {Gerstmann}}, \bibinfo {author} {\bibfnamefont
  {C.}~\bibnamefont {Gougoussis}}, \bibinfo {author} {\bibfnamefont
  {A.}~\bibnamefont {Kokalj}}, \bibinfo {author} {\bibfnamefont
  {M.}~\bibnamefont {Lazzeri}}, \bibinfo {author} {\bibfnamefont
  {L.}~\bibnamefont {Martin-Samos}}, \bibinfo {author} {\bibfnamefont
  {N.}~\bibnamefont {Marzari}}, \bibinfo {author} {\bibfnamefont
  {F.}~\bibnamefont {Mauri}}, \bibinfo {author} {\bibfnamefont
  {R.}~\bibnamefont {Mazzarello}}, \bibinfo {author} {\bibfnamefont
  {S.}~\bibnamefont {Paolini}}, \bibinfo {author} {\bibfnamefont
  {A.}~\bibnamefont {Pasquarello}}, \bibinfo {author} {\bibfnamefont
  {L.}~\bibnamefont {Paulatto}}, \bibinfo {author} {\bibfnamefont
  {C.}~\bibnamefont {Sbraccia}}, \bibinfo {author} {\bibfnamefont
  {S.}~\bibnamefont {Scandolo}}, \bibinfo {author} {\bibfnamefont
  {G.}~\bibnamefont {Sclauzero}}, \bibinfo {author} {\bibfnamefont {A.~P.}\
  \bibnamefont {Seitsonen}}, \bibinfo {author} {\bibfnamefont {A.}~\bibnamefont
  {Smogunov}}, \bibinfo {author} {\bibfnamefont {P.}~\bibnamefont {Umari}}, \
  and\ \bibinfo {author} {\bibfnamefont {R.~M.}\ \bibnamefont {Wentzcovitch}},\
  }\href {http://www.quantum-espresso.org} {\bibfield  {journal} {\bibinfo
  {journal} {Journal of Physics: Condensed Matter}\ }\textbf {\bibinfo {volume}
  {21}},\ \bibinfo {pages} {395502 (19pp)} (\bibinfo {year}
  {2009})}\BibitemShut {NoStop}%
\bibitem [{\citenamefont {Deslippe}\ \emph {et~al.}(2012)\citenamefont
  {Deslippe}, \citenamefont {Samsonidze}, \citenamefont {Strubbe},
  \citenamefont {Jain}, \citenamefont {Cohen},\ and\ \citenamefont
  {Louie}}]{Deslippe20121269}%
  \BibitemOpen
  \bibfield  {author} {\bibinfo {author} {\bibfnamefont {J.}~\bibnamefont
  {Deslippe}}, \bibinfo {author} {\bibfnamefont {G.}~\bibnamefont
  {Samsonidze}}, \bibinfo {author} {\bibfnamefont {D.~A.}\ \bibnamefont
  {Strubbe}}, \bibinfo {author} {\bibfnamefont {M.}~\bibnamefont {Jain}},
  \bibinfo {author} {\bibfnamefont {M.~L.}\ \bibnamefont {Cohen}}, \ and\
  \bibinfo {author} {\bibfnamefont {S.~G.}\ \bibnamefont {Louie}},\ }\href
  {\doibase http://dx.doi.org/10.1016/j.cpc.2011.12.006} {\bibfield  {journal}
  {\bibinfo  {journal} {Computer Physics Communications}\ }\textbf {\bibinfo
  {volume} {183}},\ \bibinfo {pages} {1269 } (\bibinfo {year}
  {2012})}\BibitemShut {NoStop}%
\bibitem [{\citenamefont {Perdew}\ \emph {et~al.}(1996)\citenamefont {Perdew},
  \citenamefont {Burke},\ and\ \citenamefont
  {Ernzerhof}}]{PhysRevLett.77.3865}%
  \BibitemOpen
  \bibfield  {author} {\bibinfo {author} {\bibfnamefont {J.~P.}\ \bibnamefont
  {Perdew}}, \bibinfo {author} {\bibfnamefont {K.}~\bibnamefont {Burke}}, \
  and\ \bibinfo {author} {\bibfnamefont {M.}~\bibnamefont {Ernzerhof}},\ }\href
  {\doibase 10.1103/PhysRevLett.77.3865} {\bibfield  {journal} {\bibinfo
  {journal} {Phys. Rev. Lett.}\ }\textbf {\bibinfo {volume} {77}},\ \bibinfo
  {pages} {3865} (\bibinfo {year} {1996})}\BibitemShut {NoStop}%
\bibitem [{\citenamefont {Monkhorst}\ and\ \citenamefont
  {Pack}(1976)}]{PhysRevB.13.5188}%
  \BibitemOpen
  \bibfield  {author} {\bibinfo {author} {\bibfnamefont {H.~J.}\ \bibnamefont
  {Monkhorst}}\ and\ \bibinfo {author} {\bibfnamefont {J.~D.}\ \bibnamefont
  {Pack}},\ }\href {\doibase 10.1103/PhysRevB.13.5188} {\bibfield  {journal}
  {\bibinfo  {journal} {Phys. Rev. B}\ }\textbf {\bibinfo {volume} {13}},\
  \bibinfo {pages} {5188} (\bibinfo {year} {1976})}\BibitemShut {NoStop}%
\bibitem [{\citenamefont {Hybertsen}\ and\ \citenamefont
  {Louie}(1986)}]{PhysRevB.34.5390}%
  \BibitemOpen
  \bibfield  {author} {\bibinfo {author} {\bibfnamefont {M.~S.}\ \bibnamefont
  {Hybertsen}}\ and\ \bibinfo {author} {\bibfnamefont {S.~G.}\ \bibnamefont
  {Louie}},\ }\href {\doibase 10.1103/PhysRevB.34.5390} {\bibfield  {journal}
  {\bibinfo  {journal} {Phys. Rev. B}\ }\textbf {\bibinfo {volume} {34}},\
  \bibinfo {pages} {5390} (\bibinfo {year} {1986})}\BibitemShut {NoStop}%
\bibitem [{\citenamefont {Rohlfing}\ and\ \citenamefont
  {Louie}(2000)}]{PhysRevB.62.4927}%
  \BibitemOpen
  \bibfield  {author} {\bibinfo {author} {\bibfnamefont {M.}~\bibnamefont
  {Rohlfing}}\ and\ \bibinfo {author} {\bibfnamefont {S.~G.}\ \bibnamefont
  {Louie}},\ }\href {\doibase 10.1103/PhysRevB.62.4927} {\bibfield  {journal}
  {\bibinfo  {journal} {Phys. Rev. B}\ }\textbf {\bibinfo {volume} {62}},\
  \bibinfo {pages} {4927} (\bibinfo {year} {2000})}\BibitemShut {NoStop}%
\bibitem [{\citenamefont {Dancoff}(1950)}]{PhysRev.78.382}%
  \BibitemOpen
  \bibfield  {author} {\bibinfo {author} {\bibfnamefont {S.~M.}\ \bibnamefont
  {Dancoff}},\ }\href {\doibase 10.1103/PhysRev.78.382} {\bibfield  {journal}
  {\bibinfo  {journal} {Phys. Rev.}\ }\textbf {\bibinfo {volume} {78}},\
  \bibinfo {pages} {382} (\bibinfo {year} {1950})}\BibitemShut {NoStop}%
\bibitem [{\citenamefont {Wang}\ \emph
  {et~al.}(2015{\natexlab{b}})\citenamefont {Wang}, \citenamefont {Kawazoe},\
  and\ \citenamefont {Geng}}]{PhysRevB.91.045433}%
  \BibitemOpen
  \bibfield  {author} {\bibinfo {author} {\bibfnamefont {V.}~\bibnamefont
  {Wang}}, \bibinfo {author} {\bibfnamefont {Y.}~\bibnamefont {Kawazoe}}, \
  and\ \bibinfo {author} {\bibfnamefont {W.~T.}\ \bibnamefont {Geng}},\ }\href
  {\doibase 10.1103/PhysRevB.91.045433} {\bibfield  {journal} {\bibinfo
  {journal} {Phys. Rev. B}\ }\textbf {\bibinfo {volume} {91}},\ \bibinfo
  {pages} {045433} (\bibinfo {year} {2015}{\natexlab{b}})}\BibitemShut
  {NoStop}%
\bibitem [{\citenamefont {Heyd}\ \emph {et~al.}(2003)\citenamefont {Heyd},
  \citenamefont {Scuseria},\ and\ \citenamefont {Ernzerhof}}]{hseref}%
  \BibitemOpen
  \bibfield  {author} {\bibinfo {author} {\bibfnamefont {J.}~\bibnamefont
  {Heyd}}, \bibinfo {author} {\bibfnamefont {G.~E.}\ \bibnamefont {Scuseria}},
  \ and\ \bibinfo {author} {\bibfnamefont {M.}~\bibnamefont {Ernzerhof}},\
  }\href {\doibase http://dx.doi.org/10.1063/1.1564060} {\bibfield  {journal}
  {\bibinfo  {journal} {The Journal of Chemical Physics}\ }\textbf {\bibinfo
  {volume} {118}},\ \bibinfo {pages} {8207} (\bibinfo {year}
  {2003})}\BibitemShut {NoStop}%
\bibitem [{\citenamefont {Rasmussen}\ \emph {et~al.}(2016)\citenamefont
  {Rasmussen}, \citenamefont {Schmidt}, \citenamefont {Winther},\ and\
  \citenamefont {Thygesen}}]{PhysRevB.94.155406}%
  \BibitemOpen
  \bibfield  {author} {\bibinfo {author} {\bibfnamefont {F.~A.}\ \bibnamefont
  {Rasmussen}}, \bibinfo {author} {\bibfnamefont {P.~S.}\ \bibnamefont
  {Schmidt}}, \bibinfo {author} {\bibfnamefont {K.~T.}\ \bibnamefont
  {Winther}}, \ and\ \bibinfo {author} {\bibfnamefont {K.~S.}\ \bibnamefont
  {Thygesen}},\ }\href {\doibase 10.1103/PhysRevB.94.155406} {\bibfield
  {journal} {\bibinfo  {journal} {Phys. Rev. B}\ }\textbf {\bibinfo {volume}
  {94}},\ \bibinfo {pages} {155406} (\bibinfo {year} {2016})}\BibitemShut
  {NoStop}%
\bibitem [{\citenamefont {Inkson}(1971)}]{INKSON197169}%
  \BibitemOpen
  \bibfield  {author} {\bibinfo {author} {\bibfnamefont {J.~C.}\ \bibnamefont
  {Inkson}},\ }\href {\doibase http://dx.doi.org/10.1016/0039-6028(71)90085-9}
  {\bibfield  {journal} {\bibinfo  {journal} {Surface Science}\ }\textbf
  {\bibinfo {volume} {28}},\ \bibinfo {pages} {69 } (\bibinfo {year}
  {1971})}\BibitemShut {NoStop}%
\bibitem [{\citenamefont {H\"user}\ \emph {et~al.}(2013)\citenamefont
  {H\"user}, \citenamefont {Olsen},\ and\ \citenamefont
  {Thygesen}}]{PhysRevB.88.245309}%
  \BibitemOpen
  \bibfield  {author} {\bibinfo {author} {\bibfnamefont {F.}~\bibnamefont
  {H\"user}}, \bibinfo {author} {\bibfnamefont {T.}~\bibnamefont {Olsen}}, \
  and\ \bibinfo {author} {\bibfnamefont {K.~S.}\ \bibnamefont {Thygesen}},\
  }\href {\doibase 10.1103/PhysRevB.88.245309} {\bibfield  {journal} {\bibinfo
  {journal} {Phys. Rev. B}\ }\textbf {\bibinfo {volume} {88}},\ \bibinfo
  {pages} {245309} (\bibinfo {year} {2013})}\BibitemShut {NoStop}%
\end{thebibliography}

%

\end{document}